\documentclass[preprint,authoryear,12pt]{elsarticle}

\usepackage{amssymb}
\usepackage{graphicx,epsfig,color,wrapfig}
\usepackage{url,amssymb}

\journal{Computational Geometry: Theory and Applications}

\begin{document}

\begin{frontmatter}

\title{On Approximating the Riemannian $1$-Center\footnote{Second revision. Source codes for reproducible research available at~\url{http://www.informationgeometry.org/RiemannMinimax/}}}

\author[labelMarc]{Marc Arnaudon} 
\ead{Marc.Arnaudon@math.univ-poitiers.fr}
\address[labelMarc]{Laboratoire de Math\'ematiques et
  Applications\\ CNRS: UMR 6086,
  Universit\'e de Poitiers\\ T\'el\'eport 2 - BP 30179\\
  F--86962 Futuroscope Chasseneuil Cedex, France}

\author[label1,label2]{Frank Nielsen\corref{cor1}}
\cortext[cor1]{Corresponding author}

\ead{Frank.Nielsen@acm.org}
\address[label1]{Ecole Polytechnique\\ Computer Science Department (LIX) \\ Palaiseau, France.}
\address[label2]{Sony Computer Science Laboratories, Inc. (FRL),\\
3-14-13 Higashi Gotanda 3F, Shinagawa-Ku,\\
Tokyo 141-0022, Japan.}

\date{January 2011, revised April 2012}

\begin{abstract}
We generalize the Euclidean $1$-center approximation algorithm of~\cite{coresets-2003} to arbitrary Riemannian geometries, and study the  corresponding convergence rate. 
We then show how to instantiate this generic algorithm to two particular settings: 
(1) the hyperbolic geometry, and 
(2) the Riemannian manifold of symmetric positive definite matrices.  
\end{abstract}


\end{frontmatter}

\def\arccosh{\mathrm{arccosh}}
\def\eqref#1{Eq.~\ref{#1}}

\newenvironment{proof}{\noindent Proof:\\ }{\qed}
\newtheorem{thm}{Theorem}
\newtheorem{lemma}{Lemma}
\newtheorem{cor}{Corollary}
\newtheorem{prop}{Proposition}
\newtheorem{remark}{Remark}
\def\Geodesic{\mathrm{Geodesic}}

\newcommand\E{\mathbb{E}}
\newcommand\EE{\mathbb{E}}
\newcommand\N{\mathbb{N}}
\newcommand\NN{\mathbb{N}}
\newcommand\R{\mathbb{R}}
\newcommand\RR{\mathbb{R}}
\renewcommand\P{\mathbb{P}}
\newcommand\PP{\mathbb{P}}
\newcommand\Q{\mathbb{Q}}
\newcommand\QQ{\mathbb{Q}}
\newcommand\TT{\mathbb{T}}
\newcommand\ZZ{\mathbb{Z}}

\newcommand{\SA}{{\mathscr A}}
\newcommand{\SC}{{\mathscr C}}
\newcommand{\SD}{{\mathscr D}}
\newcommand{\SE}{{\mathscr E}}
\newcommand{\SG}{{\mathscr G}}
\newcommand{\SF}{{\mathscr F}}
\newcommand{\SH}{{\mathscr H}}
\newcommand{\SK}{{\mathscr K}}
\newcommand{\SL}{{\mathscr L}}
\newcommand{\SM}{{\mathscr M}}
\newcommand{\SP}{{\mathscr P}}
\newcommand{\SR}{{\mathscr R}}
\newcommand{\ST}{{\mathscr T}}
\newcommand{\SV}{{\mathscr V}}
\newcommand{\SW}{{\mathscr W}}

\def\mathpal#1{\mathop{\mathchoice{\text{\rm #1}}%
   {\text{\rm #1}}{\text{\rm #1}}%
   {\text{\rm #1}}}\nolimits}
\def\Ad{\mathpal{Ad}}
\def\ad{\mathpal{ad}}
\def\Aut{\mathpal{Aut}}
\def\cotanh{\mathrm{cotanh}}
\def\End{\mathpal{End}}
\def\Hom{\mathpal{Hom}}
\def\Ric{\mathpal{Ric}}
\def\grad{\mathpal{grad}}
\def\id{\mathpal{id}}
\def\trace{\mathpal{tr}}
\def\vol{\mathpal{vol}}
\def\Ito{{\text{\rm It\^o}}}

\def\grad{\mathop{\rm grad}\nolimits}
\def\di{\displaystyle}
\def\f{\frac}
\def\a{\alpha }
\def\b{\beta }
\def\D{\Delta }
\def\d{\delta }
\def\e{\varepsilon }
\def\G{\Gamma }
\def\g{\gamma }
\def\l{\lambda }
\def\n{\nabla }
\def\Om{\Omega }
\def\om{\omega }
\def\p{\partial }
\def\s{\sigma }
\def\ceil#1{{\lceil {#1} \rceil}}

\section{Introduction and prior work}\label{Section1}

Finding the unique smallest enclosing ball (SEB) of a finite Euclidean point set $P=\{p_1, ..., p_n\}$ is a fundamental problem that was first   posed by~\cite{Sylvester:1857}.
This problem has been thoroughly investigated in the computational geometry community by~\cite{Welzl:1991} and \cite{enclosingball-2009}, where it is also known as the minimum enclosing ball (MEB),  the $1$-center problem, or the minimax optimization problem in operations research.
In practice, since computing the SEB exactly  is intractable in high dimensions, efficient approximation algorithms have been proposed.
An algorithmic breakthrough was  achieved by~\cite{OptimalCoreSet:2008} that proved the existence of a {\it core-set} $C\subseteq P$ of {\it optimal size} $|C|=\ceil {\frac{1}{\epsilon}}$ so that $r(C)\leq (1+\epsilon) r(P)$ (for any arbitrary $\epsilon>0$), where $r(S)$ denotes the radius of the SEB of $S$. Let $c(S)$ denote the ball center, i.e. the minimax center.
Since the size of the core-set depends {\it only} on the approximation precision $\epsilon$ and is {\it independent} of the dimension, core-sets have become  widely popular in high-dimensional applications such as supervised classification in machine learning (see for example, the core vector machines of~\cite{cbm-2007}).
In the work of~\cite{coresets-2003}, a fast and simple approximation algorithm is designed as follows: 

\begin{center}
\fbox{
\begin{minipage}{0.9\textwidth}
BC-ALG:\\
\begin{itemize}
\item Initialize the center $c_1\in P$, and 
\item Iteratively update the current center using the rule  
$$c_{i+1} \leftarrow c_i+\frac{f_i-c_i}{i+1},$$
 where $f_i$ denotes the farthest point of $P$ to $c_i$.
\end{itemize}
\end{minipage}
}
\end{center}

It can be proved that a $(1+\epsilon)$-approximation of the SEB is obtained after $\ceil{\frac{1}{\epsilon^2}}$ iterations, thereby showing the
existence of a core-set $C=\{f_1, f_2, ... \}$ of a size at most $\ceil{\frac{1}{\epsilon^2}}$: $r(C)\leq (1+\epsilon) r(P)$.
This simple algorithm runs in time $O(\frac{dn}{\epsilon^2})$, and has been generalized to Bregman divergences by~\cite{fittingSEB:2005} 
which include the (squared) Euclidean distance, and are the canonical distances of dually flat spaces, including the particular case of self-dual Euclidean geometry.
(Note that if we start from the optimal center $c_1=c(S)$, the first iteration yields a center $c_2$ away from $c(S)$ but  it will converge in the long run to $c(S)$.)
\cite{OptimalCoreSet:2008} proved the existence of optimal $\epsilon$-core-set of size $\ceil{\frac{1}{\epsilon}}$.
Since finding tight core-sets requires  as a black box primitive the computation of the exact smallest enclosing
balls of small-size point sets, we rather consider the Riemmanian generalization of the BC-ALG, although that even in the Euclidean case it does not deliver optimal size core-sets.

Many data-sets arising in medical imaging (see~\cite{Pennec:ETVC}) 
or in computer vision (refer to~\cite{StiefelGrassman:2011}) cannot be considered as emanating from vectorial spaces but rather as lying on curved manifolds. 
For example, the space of rotations or the space of invertible matrices are not flat, as the arithmetic average of two elements does not necessarily lie inside the space.

In this work, we extend the Euclidean BC-ALG algorithm to Riemannian geometry.
In the remainder, we assume the reader familiar with basic notions of Riemannian geometry (see~\cite{Berger-2003} for an introductory textbook) in order not to burden the paper with technical Riemannian definitions. 
However in the appendix,  we recall some specific notions which play a key role in the paper, such as geodesics, sectional curvature, injectivity radius, Alexandrov and Toponogov theorems, and cosine laws for triangles.
Furthermore, we consider probability measures instead of finite point sets\footnote{We view finite point sets as discrete uniform probability measures.}  so as to study the most general setting.

Let $M$ be a complete Riemannian manifold and $\nu$ a probability measure on $M$. 
Denote by $\rho(x,y)$ the Riemannian distance from $x$ to $y$ on $M$ that satisfies the metric axioms. 
Assume the measure support ${\rm supp}(\nu)$ is included in a geodesic ball $B(o,R)$.

Recall that if $p\in[1,\infty)$ and $f : M\to \RR$ is a measurable function then 
$$
\|f\|_{L^p(\nu)}=\left(\int_M|f(y)|^p\,\nu(dy)\right)^{1/p}
$$
and 
$$
\|f\|_{L^\infty(\nu)}=\inf\left\{a>0,\ \ \nu\left(\{y\in M,\ |f(y)|>a\}\right)=0\right\}.
$$
Let 
\begin{equation}
 \label{E9}
R_{\a,p}=\left\{
\begin{array}{ccc}
 \f12\min\left\{{\rm inj} (M), \f{\pi}{2\a}\right\}&\hbox{ if }& 1\le p<2,\\
\f12\min\left\{{\rm inj} (M), \f{\pi}{\a}\right\}&\hbox{ if } & 2\le p\le\infty
\end{array}
\right.
\end{equation}
where ${\rm inj} (M)$ is the injectivity radius (see the appendix) and $\a>0$ is such that $\a^2$ is an upper bound for the sectional curvatures in $M$ (in fact replacing $M$ by $B(o,2R)$ is sufficient, so that we can always assume that $\a>0$).
For $p\in [1,\infty]$, under the assumption that
\begin{equation}\label{E0}
R<R_{\a,p}
\end{equation}
it has been proved by~\cite{Afsari:10} that there exists a {\it unique} point $c_p$ which minimizes the following cost function
\begin{eqnarray}
H_p &:& M\to [0,\infty]\nonumber\\
&& x\mapsto \|\rho(x,\cdot)\|_{L^p(\nu)} \label{E1}
\end{eqnarray}
with $c_p\in B(o,R)$ (in fact, lying inside the closure of the convex hull of the support of~$\nu$). 

For a discrete uniform measure viewed as a  ``point cloud'' in an Euclidean space  and $p\in[1,\infty)$, we have $H_p(x)=\left(\frac{1}{n}\sum_{i=1}^n \|p_i-x\|_p^p\right)^{1/p}$, 
with $\|\cdot\|_p$ denoting the $L_p$ norm, and $H_\infty(x)$ is the distance from $x$ to its farthest point in the cloud.

In the general situation the point $c_p$ that realizes the minimum represents a notion of centrality of the measure (eg., median for $p=1$, mean for $p=2$, and minimax center for $p= \infty$).
This center is a {\it global} minimizer (not only in $B(o,R)$), and this explains why a bound for the sectional curvature is required on the whole manifold $M$ (in fact $B(o,2R)$ is sufficient, see~\cite{Afsari:10}).

Deterministic subgradient algorithms for finding $c_p$  have been considered by~\cite{Yang:10} for the median case ($p=1$). Stochastic algorithms have been investigated by~\cite{Arnaudon-Dombry-Phan-Yang:10} for the case $p\in [1,\infty)$, and a central limit theorem (CLT) for the suitably renormalized process is derived (in fact a convergence in law to a diffusion process). See also for similar algorithms minimizing other cost functions, the work of~\cite{Bonnabel-2011}.

In this work, we consider the case $p=\infty$, with $c_\infty$ denoting the minimax center. Hereafter we use  $c$ for $c_\infty$,~$H$ for~$H_\infty$ and~$R_\a$ for~$R_{\a,\infty}$.
In this case  there is no canonical deterministic algorithm which generalizes the gradient descent algorithms considered for $p\in [1,\infty)$. 
Following \eqref{E1}, $H(x)$ denotes the farthest distance from $x$ to  a point of the support of the measure ($L^\infty$-norm).   

To give an example of a Riemannian manifold, consider the space of symmetric positive definite matrices with associated Riemannian distance (see Section~\ref{Section:App}) 
\begin{equation}
 \label{E2}
\rho(P,Q)=\|\log(P^{-1}Q)\|_F=\sqrt{\sum_i\log^2\l_i}
\end{equation}
where $\l_i$ are the eigenvalues of matrix $P^{-1}Q$. 
This is a non-compact Riemannian symmetric space of nonpositive curvature (Cartan-Hadamard manifold, see~\cite{DiffGeoLang-1999}, chapter 12). 
In this context {\it any} measure $\nu$ with {\it bounded support} satisfies.~\eqref{E0} 
(since we can take $\a>0$ as small as we like), and consequently the minimizer $c$ of $H$ exists and is unique. 
We call it the $1$-center or minimax center of $\nu$. 

We generalized the BC-ALG by noticing that the iterative update is a  barycenter  of the current minimax center with the current farthest point. Thus the new position of the minimax center falls along the straight line joining these two points in Euclidean geometry.
In Riemannian geometry, the shortest path linking two points is called a geodesic (for example,   arc of a great circle for spherical geometry).
Instead of walking on a straight line, we instead walk on the geodesic to the farthest point as follows:

\begin{center}
\fbox{
\begin{minipage}{0.9\textwidth}
GEO-ALG:\\
\begin{itemize}

\item Initialize the center with $c_1\in P$, and 

\item Iteratively update the current minimax center as 
$$c_{i+1}=\Geodesic\left(c_i,f_i,\frac{1}{i+1}\right),$$
where $f_i$ denotes the farthest point of $P$ to $c_i$, and $\Geodesic(p,q,t)$ denotes the intermediate point $m$ on the geodesic passing through $p$ and $q$ such that $\rho(p,m) =t\times \rho(p,q)$.

\end{itemize}
\end{minipage}
}
\end{center}

Note that GEO-ALG generalized BC-ALG by taking the Euclidean distance $\rho(p,q)=\| p-q \|$.

The paper is organized as follows:
Section~\ref{Section:PR} gives and proves  a crucial lemma.
It is followed by the description and convergence rate analysis of our generic Riemannian algorithm in Section~\ref{Section:Analysis}.
Section~\ref{Section:App} instantiates the algorithm for the particular cases of the hyperbolic manifold and the manifold of symmetric positive definite matrices.  
 Section~\ref{Section:Concl} concludes the paper and hints at further perspectives. 
 To make the paper self-contained, the appendix recalls the fundamental notions of Riemannian geometry used throughout the paper.

\section{A key lemma}\label{Section:PR}

In this section, we assume\footnote{Any bounded measure on a Cartan-Hadamard manifold satisfies this assumption.} that ${\rm supp}(\nu)\subset B(o,R)$ and 
 $$
R<R_{\a}=\f12\min\left\{{\rm inj} (M), \f{\pi}{\a}\right\}
$$
with $\a>0$ such that $\a^2$ is an upper bound for the sectional curvatures in $M$.
The following lemma is essential for proving the convergence of the algorithm determining the minimax of $\nu$.

\begin{lemma}
 \label{P1}
There exists $\tau>0$ such that for all $x\in B(o,R)$, 
\begin{equation}
 \label{E3}
H(x)-H(c)\ge \tau\rho^2(x,c).
\end{equation}
\end{lemma}

\begin{proof} 
 The point $c$ is the center of the smallest ball which contains ${\rm supp}(\nu)$ and the radius of this ball is exactly $r^\ast:=H(c)$ (see~\cite{Afsari:09}). An immediate consequence is that~$r^\ast\le R$. Denoting by $S(c, r^\ast)$ the boundary of this ball and by $S_{c}M$ the set of unitary vectors in  $T_{c}M$, for all  $v\in S_{c}M$ there exists $y\in S(c, r^\ast)\cap {\rm supp}(\nu)$ such that 
\begin{equation}\label{E4}
\langle \overrightarrow{cy},v\rangle \le 0
\end{equation}
where $t\mapsto \g_t(c, y)$ is the geodesic from $c$ to $y$ in time one, $\dot\g_t(c, y)$ denotes derivative with respect to~$t$ and $\overrightarrow{cy}=\dot\g_0(c, y)$. Indeed, if this was not true it would contradict the minimality of $S(c, r^\ast)$ (refer to~\cite{Afsari:09}).

Now letting $t\mapsto \g_t(v)=\exp_x(tv)$ the geodesic satisfying $\dot\g_0(v)=v$, we prove~\eqref{E3} for $x=\g_t(v)$. We have 
\begin{equation}
 \label{E5}
H(\g_t(v))-H(c)\ge \rho(\g_t(v), y)-\rho(c,y)=\rho(\g_t(v), y)-r^\ast
\end{equation}
by definition of $H$.

 Then we consider a $2$-dimensional sphere $S_{\a^2}^2$ with {\it constant} curvature $\a^2$, distance function $\tilde\rho$, and in $S_{\a^2}^2$ a {\it comparison  triangle} $\tilde \g_t(\tilde v)\tilde y\tilde  c$ such that $\tilde \rho(\tilde y,\tilde  c)=~r^\ast$, $\tilde v$ is a unitary vector in $T_{\tilde c}S_{\a^2}^2$ satisfying 
\begin{equation}
 \label{E19}
\left\langle \overrightarrow{\tilde c \tilde y},\tilde v\right\rangle=\langle \overrightarrow{cy},v\rangle
\end{equation}
 Let us prove that
\begin{equation}
 \label{E6}
\tilde \rho(\tilde \g_t(\tilde v),\tilde  y)-r^\ast=\tilde \rho(\tilde \g_t(\tilde v),\tilde  y)-\rho(\tilde c,\tilde y)\ge \tau_\a \tilde\rho^2(\tilde \g_t(\tilde v),\tilde  c)
\end{equation}
for some $\tau_\a>0$
provided condition~\eqref{E4} is realized: for simplicity we will write $\tilde d=\tilde \rho(\tilde \g_t(\tilde v),\tilde  y)$. Using~\eqref{E4} and the first law of cosines (Theorem~\ref{T4} in the  appendix), we get
\begin{equation}
 \label{E10}
0\ge \cos\left(\overrightarrow{\tilde c \tilde y},\tilde v\right)=\f{\cos \left(\a\tilde d\right)-\cos \left(\a r^\ast\right)\cos (\a t)}{\sin \left(\a r^\ast\right)\sin (\a t)}
\end{equation}
which yields 
$$
\cos \left(\a\tilde d\right)- \cos \left(\a r^\ast\right)\cos (\a t)\le 0.
$$
On the other hand since $0<\a r^\ast< \f{\pi}2$ and $0\le \a\tilde d<\pi$ we have 
$$
0\le 2\sin \left(\a\tilde d\right)\cos(\a r^\ast)\sin(\a r^\ast).
$$
So we get
$$
\cos \left(\a\tilde d\right)- \cos \left(\a r^\ast\right)\cos (\a t)\le2\sin \left(\a\tilde d\right)\cos(\a r^\ast)\sin(\a r^\ast)
$$ 
which is equivalent to
\begin{eqnarray*}
&\cos^2(\a r^\ast)\cos(\a \tilde d)+\cos(\a r^\ast)\sin(\a r^\ast)\sin(\a\tilde d)-\cos(\a r^\ast)\cos(\a t)\\
&\le \sin(\a\tilde d)\cos(\a r^\ast)\sin(\a r^\ast)-\sin^2(\a r^\ast)\cos(\a \tilde d)
\end{eqnarray*}
and this in turn implies 
$$
\sin\left(\a\left(\tilde d- r^\ast\right)\right)\ge {\rm cotan}(\a r^\ast)\left(\cos\left(\a\left(\tilde d- r^\ast\right)\right)-\cos (\a t)\right)
$$
so 
$$
\liminf_{t\searrow 0}\f{\tilde \rho(\tilde \g_t(\tilde v),\tilde  y)- r^\ast}{t^2}\ge \f{\a}{2}{\rm cotan}(\a r^\ast)\ge \f{\a}{2}{\rm cotan}(\a R_{\a})
$$
uniformly in $\tilde v$. Consequently~\eqref{E6} is true for $\tilde\g_t(\tilde v)$ in a neighborhood of $\tilde c$, and since $\tilde \rho(\tilde \g_t(\tilde v),\tilde  y)-r^\ast$ does not vanish outside this neighbourhood,  by a compactness argument we prove that~\eqref{E6} is true in any compact included in $\tilde B(\tilde c, R_{\a})$, if $\tau_\a$ is sufficiently small.

To finish the proof we are left to use the Alexandrov comparison theorem (Theorem~\ref{T2} in the  appendix) with triangles $\g_t(v)yc$ and $\tilde\g_t(\tilde v)\tilde y\tilde c$ to check that the right hand side of~\eqref{E5} in $M$ is larger than the left hand side of~\eqref{E6}. This proves~\eqref{E3} in $B(c, R)\cap B(o,R)$, and for proving it in $B(o,R)$ we just have to notice that $H$ is continuous and positive on the compact set $\bar B(o,R)\backslash B(c, R)$, hence it has a positive lower bound.
\end{proof}

\section{Riemannian approximation algorithm}\label{Section3}\label{Section:Analysis}

For $x\in B(o,R)$, denote by $t\mapsto \g_t(v(x,\nu))$ a unit speed geodesic from $\g_0(v(x,\nu))=x$ to one point $y=\g_{H(x)}(v(x,\nu))$ in ${\rm supp}(\nu)$ which realizes the maximum of the distance from $x$ to ${\rm supp}(\nu)$. So $\di v=\f1{H(x)}\exp_x^{-1}(y)$. A measurable choice is always possible. Note that if $\nu$ has finite support, when there is a finite number of possibilities for $y$ it is natural to make a random uniform choice. However in a generic situation this should never happen, there should be only one choice.

We consider the following stochastic algorithm. 

\begin{center}
\fbox{
\begin{minipage}{0.9\textwidth}
RIE-ALG:\\
Fix some $\d>0$.

{\bf Step 1}
Choose a starting point $x_0\in {\rm supp}(\nu)$ and let $k=0$


{\bf Step 2} Choose a step size $t_{k+1}\in (0,\d]$ and let $x_{k+1}=\g_{t_{k+1}}(v(x_k,\nu))$, then do again step 2 with $k \leftarrow k+1$.
\end{minipage}
}
\end{center}

This algorithm generalizes the Euclidean scheme of~\cite{coresets-2003} and algorithm GEO-ALG for probability measures.
Indeed, if  
GEO-ALG is initialized with $c_{k_0}\in P$ with $k_0$ the first  
integer larger than $1/\delta$, then it suffices to take $t_k=1/k$  
for $k\ge k_0$ in RIE-ALG.

Let $a \wedge b$ denote the minimum operator $a\wedge b=\min(a,b)$.

Let
\begin{equation}\label{E21}
 R_0=\f{R_{\a}-R}{2}\wedge \f{R}{2}.
\end{equation}
 
\begin{thm}
 \label{T1}
Assume $\a,\b>0$ are such that $-\b^2$ is a lower bound and $\a^2$ an upper bound of the sectional curvatures in $M$.

If the step sizes $(t_k)_{k\ge 1}$ satisfy

\begin{eqnarray}
\d\le \f{R_0}{2}\wedge \f{2}{\b}{\rm arctanh}
\left(\tanh (\b R_0/2)\cos (\a R)\tan (\a R_0/4)\right),\label{E7} \\
\lim_{k\to\infty}t_k=0,\quad\quad\sum_{k=1}^\infty t_k=+\infty\quad\hbox{and}\quad \sum_{k=1}^\infty t_k^2<\infty.
\end{eqnarray}

then the sequence $(x_k)_{k\ge 1}$ generated by the algorithm satisfies 

\begin{equation}
 \label{E8}
\lim_{k\to\infty}\rho(x_k,c)=0.
\end{equation}

\end{thm}

\begin{remark}
 In practice $\nu$ is given and one takes any ball $B(o,R)$ which contains its support. We need the condition $R<R_{\alpha}$. One should take $R$ as small as possible for $R_0$ and then $\delta$ being not too small. The best choice is $o=c$ and $R=H(c)$ but they are not known a priori. If $\nu$ has a finite support one can take for $o$ a point of the support of $\nu$ and for $R$ the maximal distance from this point to another point of the support. It always works in a simply connected manifold of negative curvature since in this case $\alpha$ can be taken as small as we want. This is the case in our two main examples considered in Section~4, namely the hyperbolic space and the set of positive definite symmetric matrices with our specific choice of metric. Note that in this situation $R_0$ and $\delta$ can also be taken as large as we want.
\end{remark}

\begin{proof}

First we  prove that for all $r\in [R_0,R]$, if $x_k\in B(c,r)$ then $x_{k+1}\in B(c,r)$: if $\rho(x_k,c)\le R_0/2$ it is clear since $\d\le R_0/2$. If $\rho(x_k,c)\ge R_0/2$ we prove that $\rho(x_{k+1},c)\le \rho(x_k,c)$. Let $y_{k+1}=\g_{H(x_k)}(v(x_k,\nu))$: $y_{k+1}\in {\rm supp}(\nu)$ is such that $H(x_k)=\rho(x_k,y_{k+1})$; consider the triangle $c x_ky_{k+1}$. Let $a=\rho(x_k,y_{k+1})$, $b=\rho(y_{k+1},c)$ and $r=\rho(c,x_k)$, $\hat x_k$ the angle corresponding to the point $x_k$. By Alexandrov comparison theorem (in fact Corollary~\ref{C1} in the appendix)  $\hat x_k$ is smaller than the same in constant curvature $\a^2$. This together with the law of cosines in spherical geometry (Theorem~\ref{T4} in the appendix) yields 
$$
\cos \hat x_k\ge \f{\cos \a b-\cos \a r\cos \a a}{\sin\a r\sin\a a}.
$$ 
Now $r\ge R_0/2$, $b\le r^\ast$ and $a\ge r^\ast$ so 
\begin{equation}
 \label{E15}
\cos\hat x_k\ge \f{\cos \a r^\ast(1-\cos(\a R_0/2))}{\sin(\a R_0/2)}=\cos \a r^\ast\tan(\a R_0/4)\ge \cos \a R \tan(\a R_0/4).
\end{equation}
Consider now the triangle $c x_kx_{k+1}$ and let $f=\rho(c, x_{k+1})$. Recall $\rho(x_k,x_{k+1})=t_{k+1}$. Now by Toponogov theorem (Theorem~\ref{T3} in the appendix) $f$ is smaller than the same   in constant curvature $-\b^2$. This together with first law of cosines in hyperbolic geometry (Theorem~\ref{T4} in the appendix) yields  
\begin{equation}
 \label{E16}
\cosh \b f\le \cosh \b r\cosh \b t_{k+1}-\cos\hat x_k\sinh \b r\sinh \b t_{k+1}
\end{equation}
which implies by \eqref{E15}
\begin{equation}
 \label{E17}
\cosh \b f\le \cosh (\b r)\cosh \b t_{k+1}-\cos \a R \tan(\a R_0/4)\sinh (\b r)\sinh \b t_{k+1}.
\end{equation}
Let us  check that the condition on $\d$ implies that the right hand side is smaller than $\cosh \b r$: we want to prove 
$$
\cosh (\b r)(\cosh \b t_{k+1}-1)\le\cos \a R \tan(\a R_0/4)\sinh (\b r)\sinh \b t_{k+1}
$$
or equivalently
\begin{equation}
\label{E17bis}
\f{\cosh \b t_{k+1}-1}{\sinh \b t_{k+1}}\le \cos \a R \tan(\a R_0/4)\tanh (\b r).
\end{equation}
But 
$$
\f{\cosh \b t_{k+1}-1}{\sinh \b t_{k+1}}=\tanh\left(\f{\b t_{k+1}}{2}\right)
$$
and $t_{k+1}\le \d$, $r\ge R_0/2$, so that~\eqref{E17bis} is implied by
\begin{equation}
\label{E17ter}
\tanh\left(\f{\b \d}{2}\right)\le \cos \a R \tan(\a R_0/4)\tanh \left(\f{\b R_0}{2}\right).
\end{equation}
Now clearly the condition on $\d$ implies~\eqref{E17ter}.

 So we have proved that $\rho(c, x_{k+1})\le \rho(c, x_{k})$.

Then we prove that there exists $\eta>0$ such that  if $x_k\in B(c,R)\backslash B(c,R_0)$ then 
\begin{equation}\label{E20} 
\f{\cosh\left(\b\rho(c, x_{k+1})\right)}{\cosh\left(\b\rho(c, x_{k})\right)}\le 1-\eta t_{k+1}.
\end{equation}
 From~\eqref{E17}, we obtain
 
\begin{eqnarray*}
 \f{\cosh \b f}{\cosh \b r} &\le & \cosh \b t_{k+1}-\cos \a R \tan(\a R_0/4)\tanh (\b r)\sinh \b t_{k+1}\\ 
&\le  & \cosh \b t_{k+1}-\cos \a R \tan(\a R_0/4)\tanh (\b R_0)\sinh \b t_{k+1}\\
&\le & 1-2 \left(\cos \a R \tan(\a R_0/4)\tanh (\b R_0)\cosh(\b t_{k+1}/2)  \right. \\
&& \left. -\sinh(\b t_{k+1}/2) \right)\sinh(\b t_{k+1}/2)\\
&\le & 1-\left(\cos \a R \tan(\a R_0/4)\tanh (\b R_0)\cosh(\b t_{k+1}/2)-\sinh(\b t_{k+1}/2)\right)\b t_{k+1}\\
&\le & 1- (\cos \a R \tan(\a R_0/4)\tanh (\b R_0) \\
&& -\cos \a R \tan(\a R_0/4)\tanh (\b R_0/2) )\cosh(\b t_{k+1}/2)\b t_{k+1}\\
\end{eqnarray*}
where we used Eq.~\ref{E7} in the last inequality.
So 
\begin{eqnarray}
\f{\cosh \b \rho(c, x_{k+1})}{\cosh \b\rho(c, x_{k})}& \le &  1-(\cos \a R \tan(\a R_0/4)\tanh (\b R_0)  \nonumber \\   && -\cos \a R \tan(\a R_0/4)\tanh (\b R_0/2) )\b t_{k+1}\nonumber\\ \   \label{E18} 
\end{eqnarray}
and this gives Eq.~\ref{E20}.

At this stage, since $\di \sum_{k=1}^\infty t_k=\infty$,  we can conclude that there exists $k_0$ such that $\di\cosh\left(\b\rho(c, x_{k_0})\right)\le\cosh(\b R_0)$ so $x_{k_0}\in B(c,R_0)$. Moreover from the first part of the proof we have that for all $k\ge k_0$, $x_{k}\in B(c,R_0)$.

Now we use the fact that on $B(c,R_0)$, $H$ is convex and satisfies~\eqref{E3}. 
By boundedness of the Hessian of square distance to $c$ (see \cite{Yang:10} Lemma~1.1 for details), we have for $k\ge k_0$

\begin{eqnarray}
\lefteqn{\rho^2(c,x_{k+1}) \le }\nonumber \\
 && \rho^2(c,x_{k})- 2t_{k+1}\left\langle\exp_{x_k}^{-1}c,\dot\g_0(v(x_k,\nu)) \right\rangle+C\left(\f{R_{\a}+R}{2},\beta\right)t_{k+1}^2\nonumber \\ \  \label{E14}
\end{eqnarray}
with 
\begin{equation}
 \label{E11}
C(r,\beta)= 2r\b \cotanh(2\b r).
\end{equation}

Now letting $y_{k+1}=\g_{H(x_k)}(v(x_k,\nu))$ we have $H\ge \rho(\cdot ,y_{k+1})$ since $y_{k+1}\in{\rm supp}(\nu)$. We remark that $\rho^2(\cdot ,y_{k+1})$ is convex on $B(c, R_0)$
by the fact that for all $z\in B(c, R_0)$ and $y\in {\rm supp}(\nu)$, $\rho(z,y)<R_{\a}$.
Moreover we have $H(x_k)=\rho(x_k ,y_{k+1})$.
As a consequence, we get 

\begin{eqnarray*}
H(c)-H(x_k) &&\ge \rho^2(c ,y_{k+1})-\rho^2(x_k ,y_{k+1})
\\ &&\ge -2\left\langle\exp_{x_k}^{-1}c,\dot\g_0(v(x_k,\nu)) \right\rangle
\end{eqnarray*}

and this implies by Lemma~\ref{P1}
\begin{equation}
 \label{E12}
-2\left\langle\exp_{x_k}^{-1}c,\dot\g_0(v(x_k,\nu)) \right\rangle\le -\tau \rho^2(c, x_k).
\end{equation}
Plugging into~\eqref{E14} yields 
\begin{equation}
 \label{E13}
\rho^2(c,x_{k+1}) \le (1-\tau t_{k+1})\rho^2(c,x_{k})+C\left(\f{R_{\a}+R}{2},\beta\right)t_{k+1}^2.
\end{equation}
We recall from here the standard argument to prove that $\rho^2(c,x_{k})$ converges to~$0$. Let 
$$
a=\limsup_{k\to\infty}\rho^2(c,x_{k}).
$$
 Iterating~\eqref{E13}
yields for $\ell\ge 1$ 
$$
\rho^2(c,x_{k+\ell}) \le \prod_{j=1}^\ell (1-\tau t_{k+j})\rho^2(c, x_k)+C\sum_{j=1}^\ell t_{k+j}^2
$$
with $C=C\left(\f{R_{\a}+R}{2},\beta\right)$. Letting $\ell\to \infty$ and using the fact that $\di \sum_{j=1}^\infty t_{k+j}=\infty$, which implies 
$$
\prod_{j=1}^\infty (1-\tau t_{k+j})=0,
$$
we get 
$$
a\le C\sum _{j=1}^\infty t_{k+j}^2.
$$
Finally using $\sum _{j=1}^\infty t_{j}^2<\infty$ we obtain that $\lim_{k\to\infty}\sum _{j=1}^\infty t_{k+j}^2=0$, so $a=0$.

\end{proof}

\begin{remark}
In Theorem~\ref{T1}, it looks difficult to find a larger $\delta$. The choice is almost optimal to have $\rho(c, x_{k+1})\le \rho(c, x_k)$ outside $B(c, R_0)$. On the other hand \eqref{E18} yields an explicit value for $\eta$ in \eqref{E20} and this in turn can be used to find an explicit $\eta'>0$ such that 
 \begin{equation}
 \label{E22}
 \rho^2(c,x_{k+1})\le (1-\eta' t_{k+1})\rho^2(c,x_k), \quad t_{k+1}\le \d\wedge 1/\eta'.
 \end{equation}
\end{remark}

For the speed of convergence, taking $\di t_k=\f{r}{k+1}$, we proceed  as in Proposition~4.10 of~\cite{Yang:10}. We use the following lemma, borrowed from the paper of~\cite{Nedic-Bertsekas:00}:

\begin{lemma}
 \label{L1}
Let $(u_k)_{k\ge 1}$ be a sequence of nonnegative real numbers such that
$$
u_{k+1}\le \left(1-\f{\l}{k+1}\right)u_k+\f{\xi}{(k+1)^2}
$$
where $\l$ and $\xi$ are positive constants. Then 
$$
u_{k+1}\le \left\{
\begin{array}{cc}
 \f1{(k+1)^\l}\left(u_0+\f{2^\l\xi(2-\l)}{1-\l}\right)&\quad \hbox{if}\quad 0<\l<1;\\
\f{\xi\left(1+\ln(k+1)\right)}{k+1}&\quad \hbox{if}\quad \l=1;\\
\f1{(\l-1)(k+2)}\left(\xi+\f{(\l-1)u_0-\xi}{(k+2)^{\l-1}}\right)&\quad \hbox{if}\quad \l>1.\\
\end{array}
\right.
$$
\end{lemma}

\begin{prop}
 \label{P2} Choosing $\di t_k=\f{r}{k+1}$, letting $k_0$ such that for all $k\ge k_0$, $x_k\in B(c, R_0)$, 
$$
\rho^2(x_{k_0+k},c)\le \left\{
\begin{array}{cc}
 \f1{(k+1)^\l}\left(R_0^2+\f{2^\l\xi(2-\l)}{1-\l}\right)&\quad \hbox{if}\quad 0<\l<1;\\
\f{\xi\left(1+\ln(k+1)\right)}{k+1}&\quad \hbox{if}\quad \l=1;\\
\f1{(\l-1)(k+2)}\left(\xi+\f{(\l-1)R_0^2-\xi}{(k+2)^{\l-1}}\right)&\quad \hbox{if}\quad \l>1.\\
\end{array}
\right.
$$ 
where $\l=\tau r$ (with $\tau$ given in Lemma~\ref{P1}) and $\xi=r^2C\left(\f{R_{\a}+R}{2},\beta\right)$.
\end{prop}
\begin{proof}
 This is a direct consequence of lemma~\ref{L1} and inequality~\eqref{E13}, valid for $k\ge k_0$.
\end{proof}
\begin{remark}
  From the  estimate of $\eta$ given by~\eqref{E18} one can get an estimate of~$k_0$. Another possibility is to replace~$\tau$ by $\tau\wedge\eta'$ in~\eqref{E13} with~$\eta'$ defined in~\eqref{E22}. Then Proposition~\ref{P2} is valid for all $k\ge 1$ without the condition $x_k\in B(c, R_0)$.
\end{remark}

\begin{remark}
 The proof of Theorem~1 works for $R_0$ defined in~\eqref{E21}. It also works for any smaller positive value. It is better to have $R_0$ large so that $x_k$ rapidly enters the ball $B(c, R_0)$. On the other hand when $R_0$ is small  and $x_k$ is already in this ball then one can take $\tau$ close to $\f{\a}{2}{\rm cotan}(\a R_{\a})$. Again explicit estimates are possible.
\end{remark}


\section{Two case studies}\label{Section4}\label{Section:App}

In order to implement algorithm GEO-ALG (a specialization of RIE-ALG for point clouds with step sizes $t_i=\frac{1}{i+1}$), we need to describe the geodesics of the underlying manifold, and find an intermediate point $m=\Geodesic(p,q,t)$ on the geodesic passing through $p$ and $q$ such that $\rho(p,m)$=t $\rho(p,q)$.

\subsection{Hyperbolic manifold}\label{sec:hyperbolicmanifold}
A hyperbolic manifold is a complete Riemannian $d$-dimensional manifold of constant sectional curvature $-1$ that is isometric to the real hyperbolic space.
There exists several models of hyperbolic geometry.
Here, we consider the planar non-conformal Klein model where geodesics are straight lines. See~\cite{hyperbolicvoronoi:2010}. 
Although there exists no known closed-form formula for the hyperbolic centroid ($p=2$), Welzl's minimax algorithm generalizes to the Klein disk as described in~\cite{hyperbolicvoronoi:2010} to compute exactly the hyperbolic $1$-center.
The Klein Riemannian distance on the unit disk is defined by

\begin{equation}
\rho(p,q) = \arccosh \frac{1-p^{\top} q}{\sqrt{(1-p^{\top} p)(1-q^{\top} q)}}
\end{equation}
where $\arccosh(x)=\log (x+\sqrt{x^2-1})$, and the geodesic passing through $p$ and $q$ is the straight line segment

\begin{equation}
\gamma_t(p,q)=(1-t)p+tq,\ t\in[0,1].
\end{equation}

\begin{figure}
\centering

\begin{tabular}{cc}
\includegraphics[bb=0 0 606 626, width=0.35\textwidth]{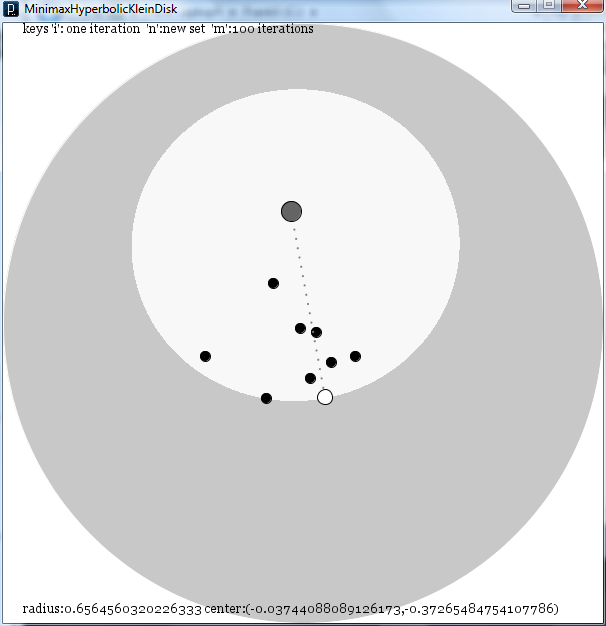} & 
\includegraphics[bb=0 0 606 626, width=0.35\textwidth]{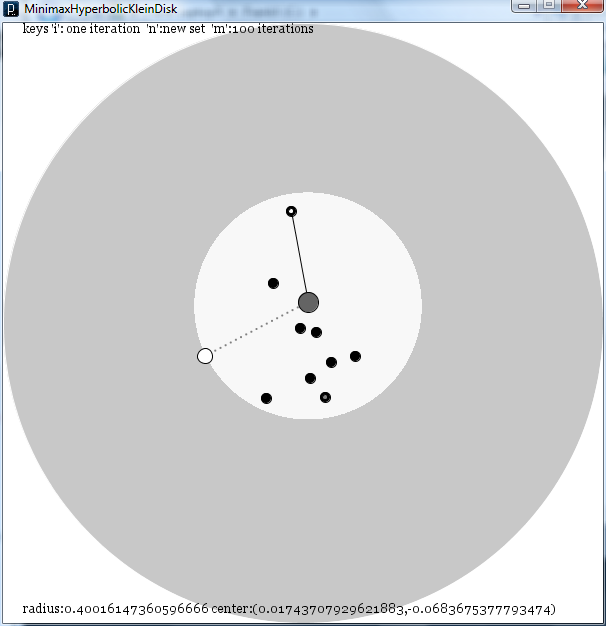} \\
Initialization & First iteration\\
\includegraphics[bb=0 0 606 626, width=0.35\textwidth]{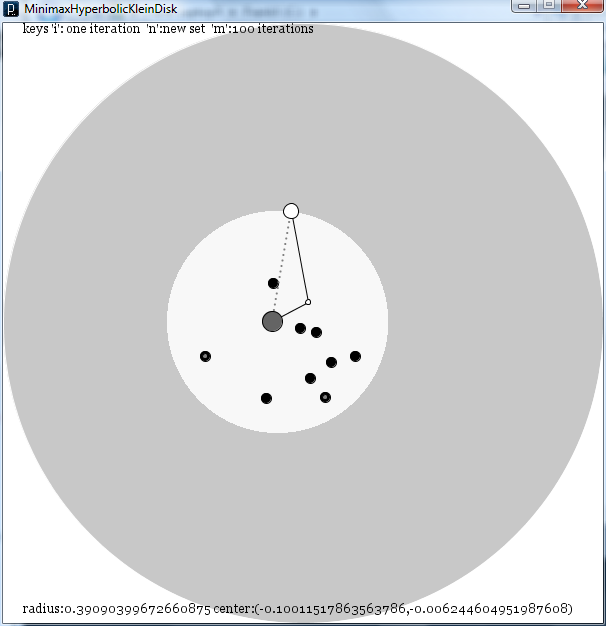} &
\includegraphics[bb=0 0 606 626, width=0.35\textwidth]{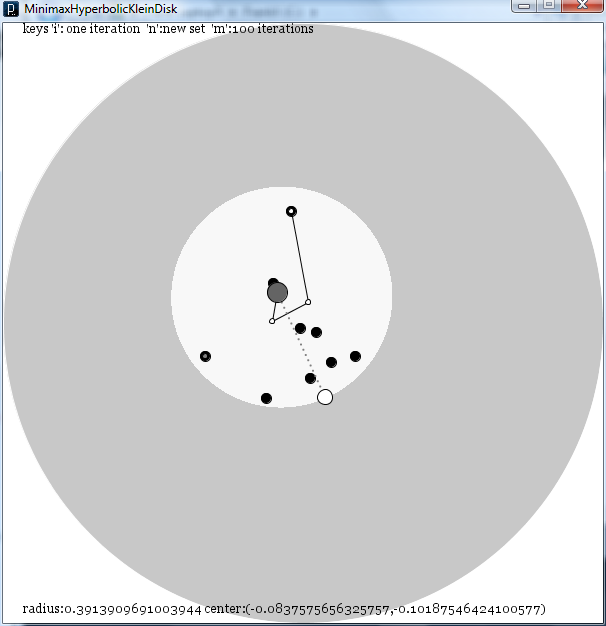} \\
Second iteration & Third iteration\\
\includegraphics[bb=0 0 606 626, width=0.35\textwidth]{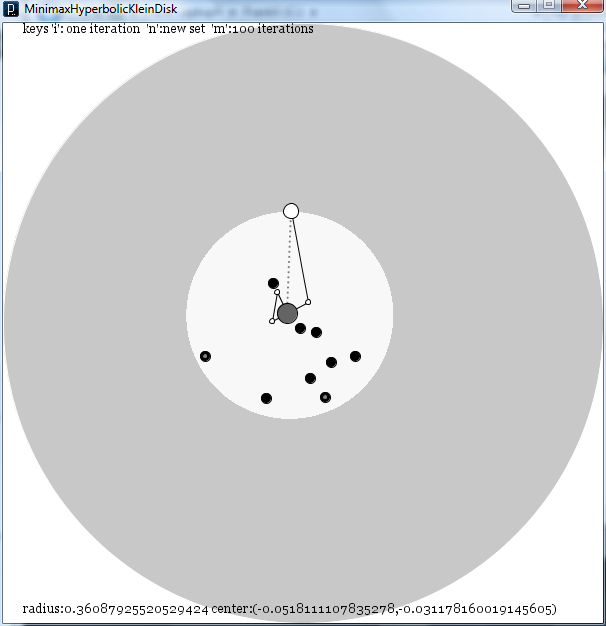} &
\includegraphics[bb=0 0 606 626, width=0.35\textwidth]{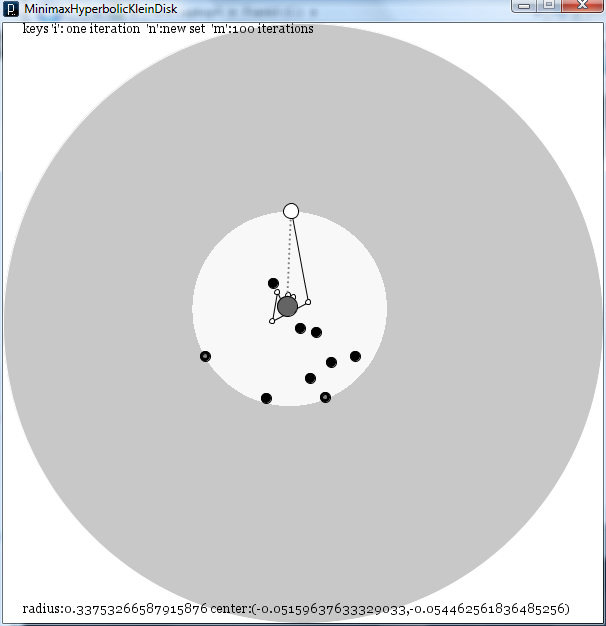}  \\
Fourth iteration & after 104 iterations
\end{tabular}
\caption{Snapshots of the GEO-ALG algorithm implemented for the hyperbolic Klein disk: 
The large black disk and the white disk denote the current   center and farthest point, respectively.
The linked path shows the trajectory of the centers as the number of iterations increase.
On-line demo available at \protect\url{http://www.informationgeometry.org/RiemannMinimax/}
 }
\label{fig:hypviz}
\end{figure}

Finding $m$ such that $\rho(p,m)$=t $\rho(p,q)$ cannot be solved in closed-form solution (except for $t=\frac{1}{2}$, see~\cite{hyperbolicvoronoi:2010}), so that we rather proceed by a bisection search algorithm on parameter $t$ up to machine precision.
Figure~\ref{fig:hypviz} shows the snapshots of our implementation in Java Processing.\footnote{\url{processing.org}}

\begin{figure}
\centering

\begin{tabular}{cc}
\includegraphics[width=0.5\textwidth]{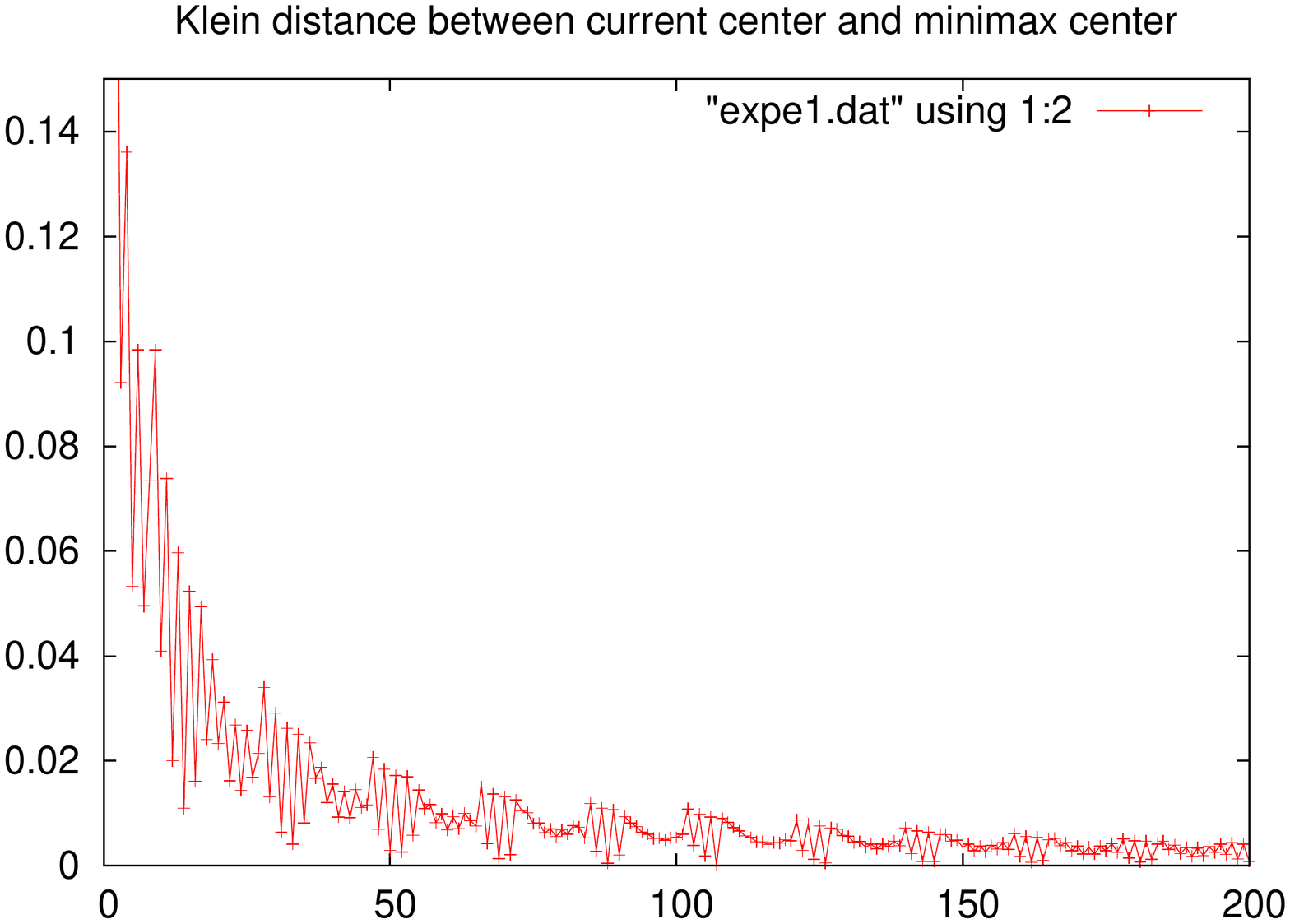}&
\includegraphics[width=0.5\textwidth]{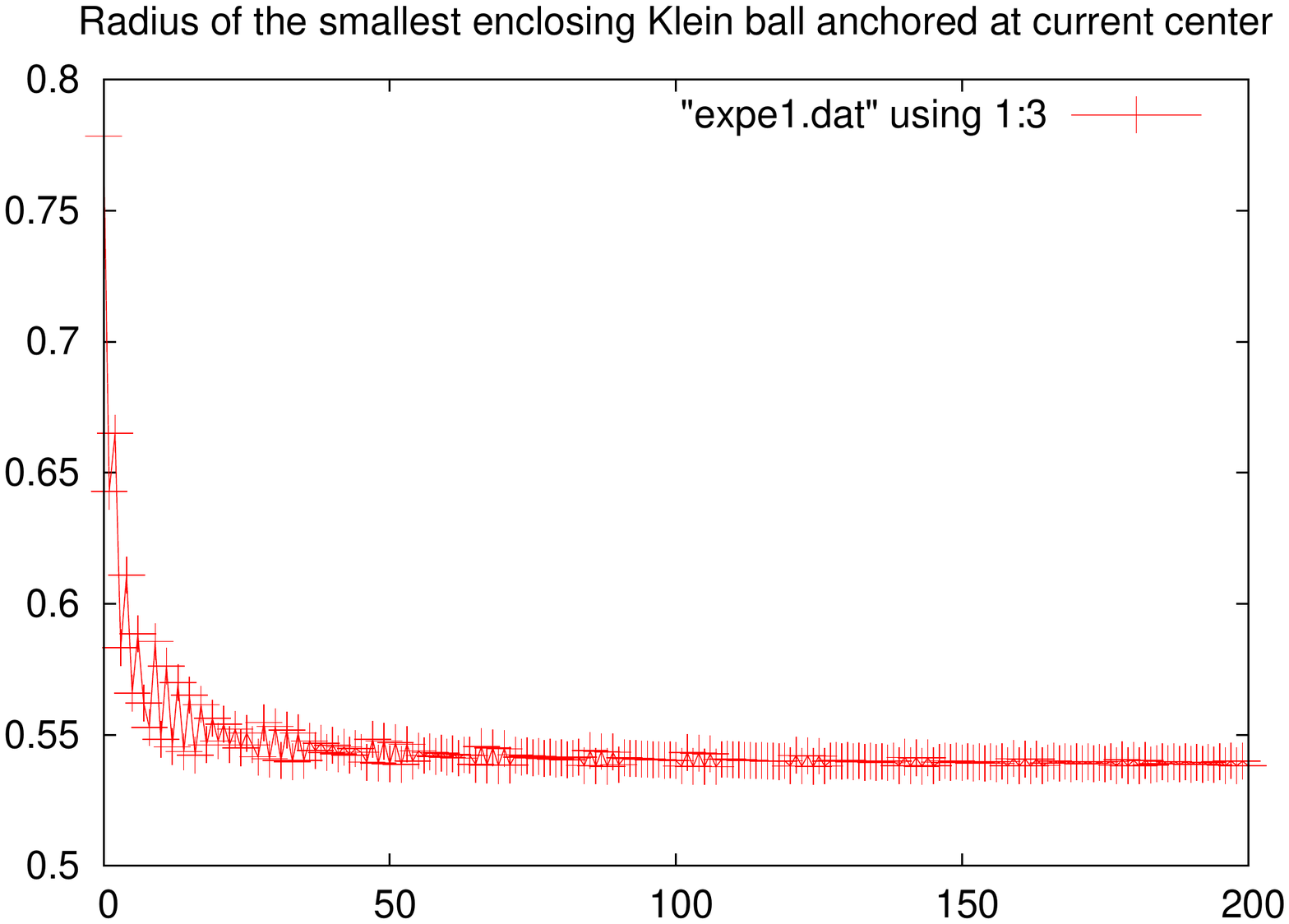}\\
(a) & (b)
\end{tabular}

\caption{Convergence rate of the GEO-ALG algorithm for the hyperbolic disk for the first 200 iterations. 
The horizontal axis denotes the number of iterations and the vertical axis (a) the relative Klein distance between the current center and the optimal $1$-center (approximated for a large number of iterations), (b) the radius of the smallest enclosing ball anchored at the current center. }

\label{fig:hypperf}

\end{figure}

Figure~\ref{fig:hypperf} plots the convergence rate of the GEO-ALG algorithm.
The code is publicly available on-line for reproducible research.

\subsection{Manifold of symmetric positive definite matrices}
\def\diag{\mathrm{diag}}

A $d\times d$ matrix $M$ with real entries is said symmetric positive definite (SPD) iff. it is symmetric ($M=M^{\top}$), and that for all $x\not =0$, $x^{\top} M x>0$.
The set of $d\times d$ SPD matrices forms a smooth manifold of dimension $\frac{d(d+1)}{2}$.
We refer to~\cite{DiffGeoLang-1999} (Chapter 12) for a description of the geometry of SPD matrices.
See also the work of~\cite{Ji} for optimization on matrix manifolds.
The geodesic linking (matrix) point $P$ to point $Q$ is given by

\begin{equation}
\gamma_t(P,Q)= P^{\frac{1}{2}} \left( P^{-\frac{1}{2}} Q P^{-\frac{1}{2}} \right)^{t} P^{\frac{1}{2}},
\end{equation}
where the matrix function $h(M)$ is computed from the singular value decomposition $M=U D V^{\top}$ (with $U$ and $V$ unitary matrices and $D=\diag(\lambda_1, ..., \lambda_d)$ a diagonal matrix of eigenvalues) as $h(M)=U \diag(h(\lambda_1), ..., h(\lambda_d)) V^{\top}$.
For example, the square root function of a matrix is computed as $M^{\frac{1}{2}}=U \diag(\sqrt{\lambda_1}, ..., \sqrt{\lambda_d}) V^{\top}$.

In this case, finding $t$ such that

\begin{equation}
\|\log (P^{-1}Q)^{t} \|_F^2 = r \| \log P^{-1}Q \|_F^2,  \label{eq:SPD}
\end{equation}
where $\|\cdot\|_F$ denotes the Fr\"obenius norm yields to $t=r$.
Indeed, consider $\lambda_1, ..., \lambda_d$ the eigenvalues of $P^{-1}Q$, then \eqref{eq:SPD} amounts to find

\begin{equation}
\sum_{i=1}^d \log^2 \lambda_i^{t} = t^2 \sum_{i=1}^d \log^2 \lambda_i = r^2 \sum_{i=1}^d \log^2 \lambda_i.
\end{equation}
That is $t=r$.

\begin{figure}
\centering

\begin{tabular}{cc}
\includegraphics[width=0.5\textwidth]{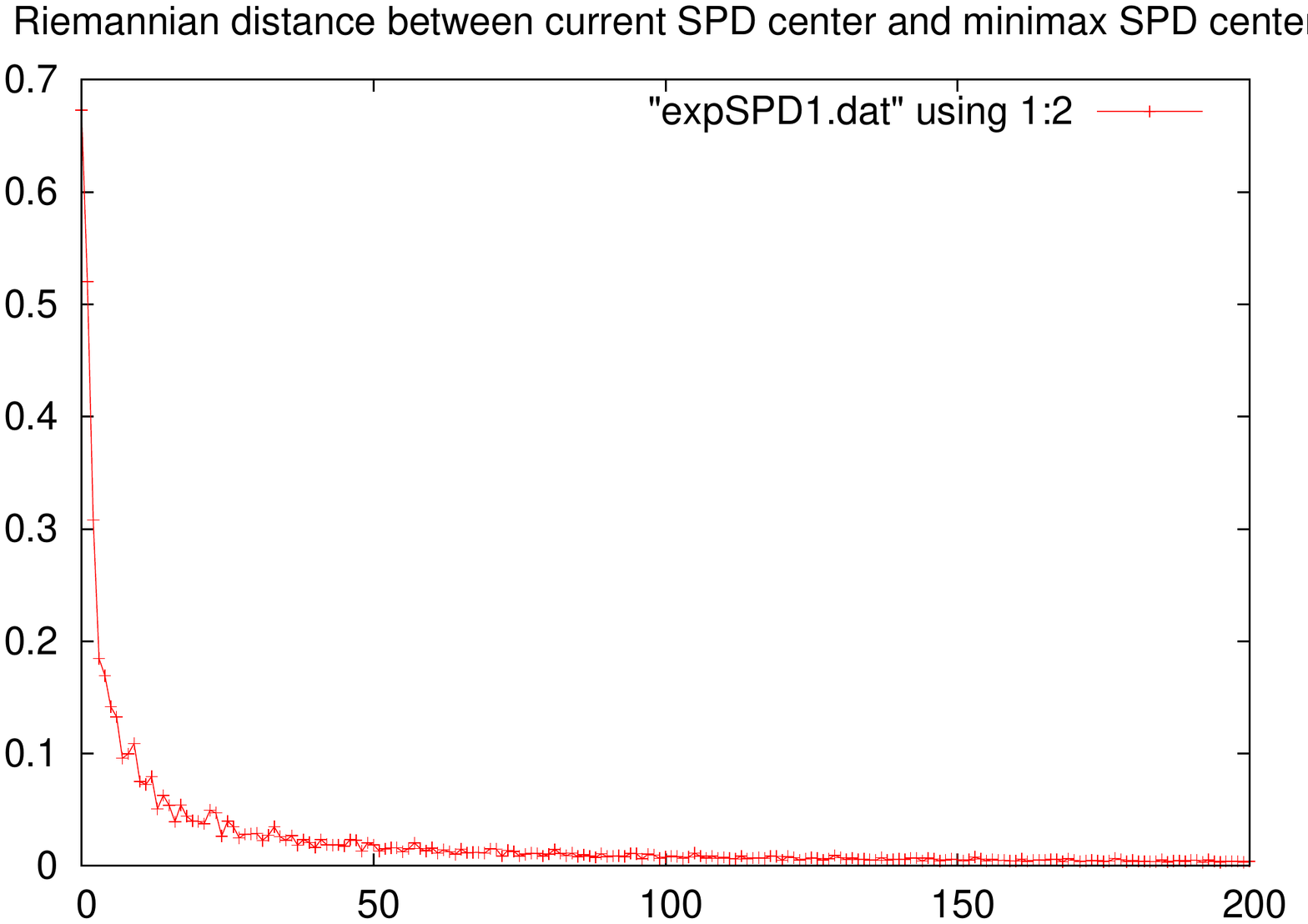}&
\includegraphics[width=0.5\textwidth]{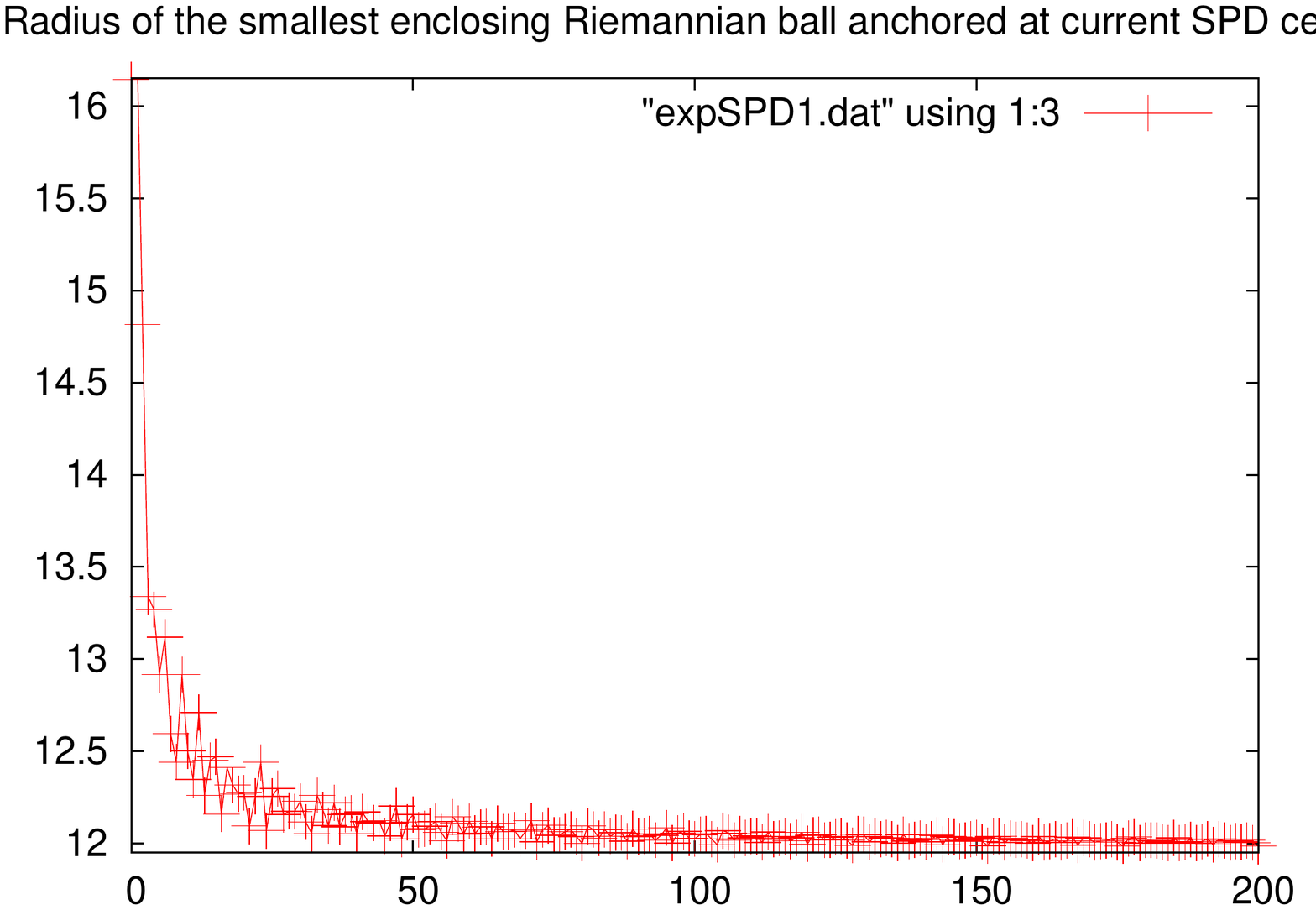}\\
(a) & (b)
\end{tabular}

\caption{Convergence rate of the GEO-ALG algorithm for the SPD Riemannian manifold (dimension $5$) for the first 200 iterations. 
The horizontal axis denotes the number of iterations $i$ and the vertical axis (a) the relative Riemannian distance between the current center $c_i$ and the optimal $1$-center $c^*$ ($\frac{\rho(c*,c_i)}{r^*}{}$, where $\rho*$ and $r^*$ are approximated for a large number of iterations), (b) the radius $r_i$ of the smallest enclosing SPD ball anchored at the current center. }

\label{fig:spdstats}

\end{figure}

Figure~\ref{fig:spdstats} displays the plots of the convergence rate of the algorithm for the SPD manifold.

\section{Concluding remarks and discussion}\label{Section:Concl}

We described a generalization of the  $1$-center algorithm of~\cite{coresets-2003} to arbitrary Riemannian geometry, and proved the convergence under mild assumptions. This proves the existence of Riemannian core-sets for optimization. 
This $1$-center building block can be used for $k$-center clustering.
Furthermore, the algorithm can be straightforwardly extended to sets of geodesic balls.

An open-source source code implementation in Java\texttrademark{} for reproducible research is available on-line at\\

\centerline{
\url{http://www.informationgeometry.org/RiemannMinimax/}
}

\section*{Acknowledgements}
The authors would like to thank the anonymous reviewers for their valuable comments and suggestions.
FN (5793b870) thanks Mr. Prasenjit Saha for discussions related to this topic, and gratefully acknowledge financial support from French funding agency ANR (GAIA 07-BLAN-0328-01) and Sony Computer Science Laboratories, Inc.

\section{Appendix: Some notions of Riemannian geometry}\label{Section:Appendix}
In this section, we recall some basic notions of Riemannian geometry used throughout the paper.
For a complete presentation, we refer to~\cite{Riem-geom:75}.

We let $M$ be a Riemannian manifold and $\langle\cdot,\cdot\rangle$ the Riemannian metric, which is a definite  positive bilinear form on each tangent space $T_xM$, and depends smoothly on $x$. The associated norm in $T_xM$ will be denoted by $\|\cdot\|$: $\|u\|=\langle u,u\rangle^{1/2}$. We denote by $\rho(x,y)$ the distance between two points on the manifold $M$:
$$
\rho(x,y)=\inf\left\{\int_0^1\|\dot\varphi(t)\|\,\mathrm{d}t,\ \ \varphi\in C^1([0,1],M),\ \ \varphi(0)=x,\ \ \varphi(1)=y\right\}.
$$

A {\it geodesic} in $M$ is a smooth path which locally minimizes the distance between two points. In general such a curve does not minimize it globally. 
However it is true in all the sets we are considering in this paper. 
Given a vector $v\in TM$ with  base point $x$, there is a unique geodesic started at $x$ with speed $v$ at time $0$. 
It is denoted by $t\mapsto \exp_x (tv)$ or compactly by $t\mapsto \g_t(v)$. 
It depends smoothly on $v$ but it has in general finite lifetime. A geodesic defined on a time interval $[a,b]$ is said to be {\it minimal} if it minimizes the distance from the image of $a$ to the image of $b$.  If the manifold is complete, taking $x,y\in M$, there exists a minimal geodesic from $x$ to $y$ in time $1$. In all the scenarii we are considering in this paper, the minimal geodesic is unique and depends smoothly on $x$ and $y$,  and we denote it by $\g_\cdot(x,y) : [0,1]\to M$,  $t\mapsto \g_t(x,y)$ with the conditions $\g_0(x,y)=x$ and $\g_1(x,y)=y$. A subset $U$ of $M$ is said to be {\it convex} if for any $x,y\in U$, there exists a unique minimal geodesic $\g_\cdot(x,y)$ in $M$ from $x$ to $y$, this geodesic fully lies in $U$ and depends smoothly on $x,y,t$.

The {\it injectivity radius} of $M$, denoted by ${\rm inj}(M)$, is the largest $r>0$ such that for all $x\in M$,
the map $\exp_x$ restricted to the open ball in $T_xM$ centered at $0$ with radius~$r$ is an embedding.

Given $x\in M$, $u,v$ two non collinear vectors in $T_xM$, the {\it sectional curvature} ${\rm Sect}(u,v)=K$ is a number which gives information on how the geodesics issued from $x$  behave near $x$. More precisely the image by $\exp_x$ of the circle centered at $0$ of radius $r>0$ in ${\rm Span}(u,v)$ has length 
$$
2\pi S_K(r)+o(r^3)\quad \hbox{as}\quad r\to 0
$$
with 
$$
S_K(r)=\left\{
\begin{array}{ccc}
 \f{\sin(\sqrt{K}r)}{\sqrt{K}}&\hbox{ if }&K>0,\\
r&\hbox{ if }&K=0,\\
\f{\sinh(\sqrt{-K}r)}{\sqrt{-K}}&\hbox{ if }&K<0.
\end{array}
\right.
$$

For instance, if $K>0$, $\di \exp_x({\rm Span}(u,v))$ is near $x$ approximatively a $2$-dimensional sphere with radius $\di\f{1}{\sqrt K}$. In fact, if $M$ is simply connected and all the sectional curvatures are equal to the same $K>0$, then $M$ is a $d$-dimensional sphere with radius $\di\f{1}{\sqrt K}$, where $d$ is the dimension of $M$. If $M$ is simply connected and all the sectional curvatures are equal to the same $K<0$, we say that $M$ is a $d$-dimensional hyperbolic space with curvature~$K$.

An upper bound (resp. lower bound) of sectional curvatures is a number $a$ such that for all non collinear $u,v$ in the same tangent space, ${\rm Sect}(u,v)\le a$ (resp. ${\rm Sect}(u,v)\ge a$). 
In the paper,  we  used a positive upper bound $\a^2$ and a negative lower bound $-\b^2$, $\a,\b>0$.

The existence of the upper bound $\a^2$  for sectional curvatures makes possible to compare geodesic triangles, by {\it Alexandrov} theorem (see \cite{Comp-th:06}). 
\begin{thm}
\label{T2}
 Let $x_1,x_2,x_3\in M$ satisfy $x_1\not=x_2$, $x_1\not=x_3$ and $$ \rho(x_1,x_2)+\rho(x_2,x_3)+\rho(x_3,x_1)<2\min\left\{{\rm inj}M, \f{\pi}{\a}\right\}$$
where $\a>0$ is such that $\a^2$ is an upper bound of sectional curvatures. Let the minimizing geodesic from $x_1$ to $x_2$ and the minimizing geodesic from $x_1$ to $x_3$ make an angle $\theta$ at $x_1$. 
Denoting by $S_{\a^2}^2$ the $2$-dimensional sphere of constant curvature $\a^2$ (hence of radius $1/\a$) and $\tilde \rho$ the distance in $S_{\a^2}^2$, we consider points $\tilde x_1,\tilde x_2, \tilde x_3\in S_{\a^2}^2$ such that $\rho(x_1,x_2)=\tilde \rho(\tilde x_1,\tilde x_2)$, $\rho(x_1,x_3)=\tilde \rho(\tilde x_1,\tilde x_3)$. Assume that the minimizing geodesic from $\tilde x_1$ to $\tilde x_2$ and the minimizing geodesic from $\tilde x_1$ to $\tilde x_3$  also make an angle $\theta$ at $\tilde x_1$.

Then we have $\rho(x_2,x_3)\ge \tilde \rho(\tilde x_2,\tilde x_3)$.
\end{thm}
Instead of prescribing the angle in the comparison triangle in the sphere, it is possible to prescribe the third distance:
\begin{cor}
 \label{C1}
The assumption are the same as in Theorem~\ref{T2} except that we assume that  $\rho(x_2,x_3)= \tilde \rho(\tilde x_2,\tilde x_3)$ (all the distances are equal), but the minimizing geodesic from $\tilde x_1$ to $\tilde x_2$ and the minimizing geodesic from $\tilde x_1$ to $\tilde x_3$ now make an angle $\tilde\theta$ at $\tilde x_1$.

Then we have $\tilde \theta\ge \theta$.
\end{cor}

There also exists a comparison result in the other direction, called Topogonov's theorem. 

\begin{thm}
\label{T3}
Assume $\b>0$ is such that $-\b^2$ is a lower bound for sectional curvatures in~$M$.
Let $x_1,x_2,x_3\in M$ satisfy $x_1\not=x_2$, $x_1\not=x_3$. Let the minimizing geodesic from $x_1$ to $x_2$ and the minimizing geodesic from $x_1$ to $x_3$ make an angle $\theta$ at $x_1$. 
Denoting  by $H_{-\b^2}^2$ the hyperbolic $2$-dimensional space of constant curvature $-\b^2$ and $\tilde \rho$ the distance in $H_{-\b^2}^2$, we consider points $\tilde x_1,\tilde x_2, \tilde x_3\in H_{-\b^2}^2$ such that $\rho(x_1,x_2)=\tilde \rho(\tilde x_1,\tilde x_2)$, $\rho(x_1,x_3)=\tilde \rho(\tilde x_1,\tilde x_3)$. Assume that the minimizing geodesic from $\tilde x_1$ to $\tilde x_2$ and the minimizing geodesic from $\tilde x_1$ to $\tilde x_3$ also make  an angle $\theta$ at $\tilde x_1$.

Then we have $\rho(x_2,x_3)\le \tilde \rho(\tilde x_2,\tilde x_3)$.
\end{thm}

Triangles in the sphere $S_{\a^2}^2$ and in the hyperbolic space $H_{-\b^2}^2$ have explicit relations between distance and angles as we will see below. This combined with Theorems~\ref{T2} and~\ref{T3} and Corollary~\ref{C1} allow to find related bounds in $M$, which are intensively used in our proofs. 

In this paper, we only use the {\it first law of cosines} in $S_{\a^2}^2$ and in $H_{-\b^2}^2$ (see e.g., the paper of~\cite{Law-Cos:94} Theorem~2.5.3 and Theorem~3.5.3).
\begin{thm}
 \label{T4}
If $\theta_1,\theta_2,\theta_3$ are the angles of a triangle in $S_{\a^2}^2$ and $x_1,x_2,x_3$ are the lengths of the opposite sides, then
$$
\cos \theta_3=\f{\cos (\a x_3)-\cos (\a x_1)\cos (\a x_2)}{\sin (\a x_1)\sin (\a x_2)}.
$$
If $\theta_1,\theta_2,\theta_3$ are the angles of a triangle in $H_{-\b^2}^2$ and $x_1,x_2,x_3$ are the lengths of the opposite sides, then
$$
\cos \theta_3=\f{\cosh (\b x_1)\cosh (\b x_2)-\cosh (\b x_3)}{\sinh (\b x_1)\sinh (\b x_2)}.
$$
\end{thm}


\end{document}